\documentstyle[version2,aps,latexsym,twocolumn]{revtex}
\begin{document}
\widetext
\begin{title}
How generic are null spacetime singularities?
\end{title}
\author{Amos Ori}
\begin{instit}
Department of Physics, Technion - Israel Institute of Technology,
32000 Haifa, Israel.
\end{instit}
\author{\'Eanna \'E. Flanagan}
\begin{instit}
Enrico Fermi Institute, University of Chicago, Chicago, IL 60637-1433.
\end{instit}
\input epsf.tex
\def\plotoneNew#1{\begin{center} 
    \epsfxsize=0.5\textwidth \epsfbox{#1} \mbox{} \end{center}}

\begin{abstract}

The spacetime singularities inside realistic black holes are sometimes
thought to be spacelike and strong, since there is a generic class of
solutions (BKL) to Einstein's equations with these properties.  We show
that null, weak singularities are also generic, in the following
sense: there is a class of vacuum solutions containing null, weak
singularities, depending on 8 arbitrary (up to some inequalities)
analytic initial functions of 3 spatial coordinates.  Since 8
arbitrary functions are needed (in the gauge used here) to span the
generic solution, this class can be regarded as generic.
\end{abstract}

\narrowtext 

\twocolumn

One of the most fascinating outcomes of general relativity is the
observation that the most fundamental concept in physics --- the 
fabric
of space and time --- may become singular in certain circumstances.  A
series of singularity theorems \cite{singularity_theorems} imply
that spacetime singularities are expected to develop inside black holes.  
The
observational evidence at present is that black holes do exist in the
Universe.  The formation of spacetime singularities in the real world
is thus almost inevitable.  However, the singularity theorems tell us
almost nothing about the nature and location of these singularities.
Despite a variety of investigations, there is today still no consensus
on the structure of singularities inside realistic black holes.

At issue are the following features of singularities: their location,
causal character (spacelike, timelike or null), and, most importantly,
their strength.  We use here Tipler's terminology \cite{Tipler} for
weak and strong singularities.  In typical situations, if the spacetime can
be extended through the singular hypersurface so that the metric
tensor is $C^0$ and non-degenerate, then the singularity is weak
\cite{Tipler}.  The strength of the singularity has far-reaching
physical consequences.  A physical object which moves towards a strong
curvature singularity will be completely torn apart by the diverging
tidal force, which causes unbounded tidal distortion. On the other
hand, if the singularity is weak, the total tidal distortion may be
finite (and even arbitrarily small), so that physical observers may
possibly not be destroyed by the singularity
\cite{Tipler,Ori_charged}.

The main difficulty in determining the structure of black hole
singularities is that the celebrated exact black hole solutions
(the Kerr-Newman family) do not give a 
realistic description of the geometry
inside the horizon, although they do describe well the region outside.
This is because the well known no-hair property of black holes ---
that arbitrary initial perturbations are harmlessly radiated away and
do not qualitatively change the spacetime structure  --- only applies
to the {\it exterior} geometry.  The geometry inside the black hole
(near the singularity and/or the Cauchy horizon) is
unstable to initial small perturbations \cite{Novikov,Hartle},
and consequently we must go beyond the
classic exact solutions to understand realistic black hole interiors.  To
determine the structure of generic singularities, it is necessary
to take initial data corresponding to the classic black hole
solutions, make generic small perturbations
 to the initial data, and evolve
forward in time to determine the nature of 
the resulting singularity.  For
this purpose a linear evolution of the perturbations may be insufficient 
 --- the real question is what happens in full nonlinear general 
relativity.

The simplest black-hole solution, the Schwarzschild solution, contains 
a central singularity which is spacelike and strong.  For many years,
this Schwarzschild singularity was regarded as the archetype for a
spacetime singularity. Although this particular
type of singularity is known today to be unstable to deviations from
spherical symmetry (and hence
unrealistic) \cite{Novikov1}, another type of a strong spacelike
singularity, the so-called BKL singularity \cite{BKL}, is believed to
be generic (below we shall further explain and discuss the concept of
genericity). Since the BKL singularity is so far the {\it only} known
type of generic singularity, in the last two decades it has been
widely believed that the final state of a realistic gravitational
collapse must be the strong, spacelike, oscillatory, BKL singularity.

Recently, there have been a variety of indications that a spacetime
singularity of a completely different type actually forms inside
realistic (rotating) black holes. In particular, this singularity is
{\it null} and {\it weak}, rather than spacelike and strong.  The
first evidence for this new picture came from the mass-inflation 
model
\cite{Poisson_Israel,Ori_charged} --- a toy-model in which the Kerr
background is modeled by the spherically-symmetric Reissner-
Nordstrom
solution, and the gravitational perturbations are modeled in terms of
two crossflowing null fluids.  Later more realistic analyses replaced
the null fluids by a spherically symmetric scalar field \cite{Scalar}.
More direct evidence came from a
non-linear perturbation analysis of the inner structure of rotating
black holes \cite{Ori_Kerr}.  Both the mass-inflation models and the
nonlinear perturbation analysis of Kerr strongly suggest that a null,
weak, scalar-curvature singularity develops at the inner horizon of
the background geometry. (See also an earlier model by Hiscock
\cite{Hiscock}.)

Despite the above compelling evidence, there still is a debate
concerning the nature of generic black-hole singularities.  It is
sometimes argued that the Einstein equations, due to their
non-linearity, do not allow generic solutions with null curvature
singularities \cite{Yurtsever}. According to this argument, the
non-linearity, combined with the diverging curvature, immediately
catalyzes the transformation of the null curvature singularity into a
strong spacelike one --- presumably the BKL singularity
\cite{caveat2}.  A similar argument was also given, some  
time ago, by Chandrasekhar and Hartle \cite{Hartle}.  According to
this point of view, the results of the non-linear perturbation
analysis of Kerr are to be interpreted as an artifact of the
perturbative approach used \cite{Brady_old} (and the mass-inflation
model is a toy-model, after all).  This objection clearly marks the
need for a more rigorous, non-perturbative, mathematical analysis, to
show that a generic null weak singularity is consistent with Einstein's
equations.

Recently, Brady and Chambers showed that a null singularity could be
consistent with the constraint section of Einstein's equations
formulated on null hypersurfaces \cite{Brady_Chambers}.  However,
their result does not completely resolve the 
above issue.  The hypothesis raised in
Ref.~\cite{Yurtsever}, according to which nonlinear effects will
immediately transform the singular initial data into a spacelike
singularity, is not necessarily inconsistent with the analysis of
Ref.~\cite{Brady_Chambers}.  It is possible that a spacelike
singularity could form just at the intersection point of the two
characteristic null hypersurfaces considered in
Ref.~\cite{Brady_Chambers}.  It is primarily the {\it evolution}
equations which will determine whether singular initial data will
evolve into a null singularity or into a spacelike
one.

The purpose of this paper is to present a new mathematical
analysis which addresses the above question.  Our analysis shows that
(i) the vacuum Einstein equations (both the constraint and evolution
equations) admit solutions with a null weak singularity, and (ii) the
class of such singular solutions is so large that it depends on the
maximum possible number of independent functional degrees of 
freedom.
We will call such classes of solutions {\it functionally generic} (see
below).  Therefore any attempt to argue, on local
grounds, that a null weak singularity is necessarily inconsistent with
the non-linearities of Einstein's equations, must be false.  In the
present letter we outline this analysis and present the main results;
a full account of this work is given in Ref.~\cite{Flanagan_Ori}.

Let us first explain what we mean by ``degrees of freedom'' and
``functionally generic''.  Suppose that $\psi$ is some field on a 3+1
dimensional spacetime, which may be a multi-component field.  
Suppose
that initial data for $\psi$ are specified on some spacelike
hypersurface $S$.  We shall say that $\psi$ has $k$ ``degrees of
freedom'' if $k$ is the number of initial functions (i.e. functions of
the 3 spacelike coordinates parameterizing $S$) which need to be
specified on $S$ in order to uniquely determine inside $D^+(S)$ the
solution to the field equations satisfied by $\psi$ \cite{note2}.  The
number $k$ depends on the type of field, and also possibly on the
gauge condition used if there is gauge freedom.  For example, for a
scalar field $k=2$, because one needs to specify both $\psi$ and
${\dot \psi}$ on $S$.  For the gravitational field, it is well known
that there are $2\times 2=4$ {\it inherent} degrees of freedom. The
{\it actual} number $k$, however, is 4 plus the number of unfixed
gauge degrees of freedom, which depends on the specific gauge
conditions used.  In the gauge we use, we find that $k=8$ (see below).

We shall say that a class of solutions to the field equations is {\it
functionally generic}, if this class depends on $k$ arbitrary
functions of three independent variables \cite{note5}. This concept of 
genericity
is basically the same as that used by BKL \cite{BKL}.  The motivation
behind this definition is obvious: Suppose that a given particular
solution admits some specific feature (e.g.~a singularity of some
type). Obviously, in order for this feature to be stable to small (but
generic) perturbations in the initial data, it is necessary that the
class of solutions satisfying this feature should depend on $k$
arbitrary functions.  Functional genericity is thus a necessary
condition for stability, and is also necessary in order that there be
an open set in the space of solutions with the desired feature, in any
reasonable topology on the space of solutions \cite{note3}.

As we mentioned above, our result is  
a mathematical demonstration of the existence of a functionally generic 
null 
weak singularity.  More specifically, we prove that there exists a
class of solutions $(M,g)$ 
to the vacuum Einstein equations, which 
all admit a weak curvature 
singularity on a null hypersurface, and which depend on $k=8$ 
(see below) arbitrary analytic functions of three independent 
variables. (In Ref.~\cite{Flanagan_Ori} we shall give a more precise 
formulation of this statement.)  
The singularities may
also be characterized by the fact that the manifold may be
extended through the null surface to an
analytic manifold $(M^\prime,g^\prime)$ where the 
metric $g^\prime$ is analytic everywhere except on the null
surface where it is only $C^0$. Our construction is {\it local} in the 
sense that the manifolds we
construct are extendible (in directions away from the null
singularity); roughly speaking they can be thought of as open regions
in a more complete spacetime, part of whose boundary consists of the
singular null hypersurface.  We do {\it not} prove that null weak
singularities arise in the maximal Cauchy evolution of any
asymptotically flat, smooth initial data set.  
The spacetimes we construct are of the form $D^+(\Sigma)$, where 
$\Sigma$ is an open region in an analytic initial data set.
The curvature singularity is already present on the boundary of $\Sigma$ in the
initial data, in the sense that curvature invariants blow up along
incomplete geodesics. We emphasize that we do {\it not} view $\Sigma$ as a 
physically-acceptable initial hypersurface; rather, the initial 
hypersurface $\Sigma$ is merely a mathematical tool that we use to 
construct and parameterize the desired class of vacuum solutions.

We shall first demonstrate the main idea behind our 
mathematical construction by applying it to a simpler problem 
--- a scalar field. Consider, as an example, a real scalar 
field $\phi$ in flat spacetime, satisfying the non-linear 
field equation 
\begin{equation}
\phi _{,\alpha }^{\,\,,\alpha }=V(\phi )
\label{scalareqn}
\end{equation}
where $V(\phi )$ is some non-linear analytic function. (We 
add this non-linear piece in order to obtain a closer analogy with
the non-linear gravitational case.)  In order to show 
that this field admits a functionally generic null singularity, we 
proceed 
as follows: Let $x,y,u,v$ be the standard, double-null, 
Minkowski coordinates (i.e.~such that
$ds^2=-4 du dv+dx^2+dy^2$ ). 
Equation (\ref{scalareqn}) reads
\begin{equation}
\phi _{,uv}=\phi _{,a}^{,a}-V(\phi )
\label{scalareqn1}
\end{equation}
where here and below the indices $a,b,...$ run over the 
coordinates $x$ and $y$.  We now define 
\begin{equation}
w\equiv v^{1/ n}
\label{eqn20}
\end{equation}
for some odd integer $n\ge 3$.  We also define
\begin{equation}
t\equiv w+u\;\;,\;\; z\equiv w-u.
\label{eqn45}
\end{equation}
Re-expressing the field equation (\ref{scalareqn1}) in terms of $t$
and $z$, we obtain
\begin{equation}
\phi _{,tt}=\phi _{,zz}+n \left[ {(z+t)/ 2}
 \right]^{n-1}\left[ {\phi _{,a}^{,a}-V(\phi )} \right].
\label{eqn50}
\end{equation}

Let $M^0$ denote some neighborhood of the origin ($x=y=z=t=0$) with
compact closure, and let $S^+$ be the intersection of the hypersurface
t=0 with $M^0$ (see Fig. 1). Let $f_1(x,y,z)$ and $f_2(x,y,z)$ be two
analytic functions of their arguments, defined on $S^+$.  For any such
pair of functions, there exists a neighborhood $M^+\subseteq M^0$ of
$S^+$, and a unique analytic solution $\phi
(x,y,z,t)$ to the field equation (\ref{eqn50}) in $M^+$, such that on
$S^+$, $\phi=f_1$ and $\phi_{,t}=f_2$.  This follows directly from the
Cauchy-Kowalewski theorem \cite{Wald}, in view of the form of
Eq.~(\ref{eqn50}).  Let us denote the intersection of $M^+$ with the
null hypersurface $v=0$ by $N^+$.  Recall that $N^+$ includes a
neighborhood of the origin in the hypersurface $v=0$.

Returning now to the original independent variables ($u,v$), 
we find that $\phi (x,y,u,v)$ is continuous throughout  $M^+$. 
We now focus attention on the section $v<0$, $t \ge 0$ of $M^+$, 
which we denote by  $M$. Since the transformation from 
($z,t$) to ($u,v$) is analytic as long as $v\ne 0$, we find 
that $\phi (x,y,u,v)$ is analytic throughout  $M$.  However, 
$\phi$ will generally fail to be smooth at  $v=0$:  $\phi 
_{,v}=(1/ n)v^{1/ n-1}\phi _{,w}$ will diverge at  $v=0$ as 
long as $\phi _{,w}\ne 0$ there.  We assume that at the 
origin, $\partial f_1/ \partial z\ne \pm f_2$.  This ensures that at 
least in 
some neighborhood of the origin, both $\phi _{,w}$ and $\phi 
_{,u}$ are nonzero.  Let  $N$ be the intersection of that 
neighborhood with the section $t\ge 0$ of  $N^+$.  We find that 
$\phi _{,v}$ diverges on  $N$.  Moreover, the invariant $\phi 
_{,\alpha }\phi ^{,\alpha} $ diverges on  $N$ too [it is 
dominated by $(1/ n)v^{-1+1/n}\phi _{,u}\phi _{,w}$ ].  $N$ 
is thus a singular null hypersurface.

{\vskip -1.2cm}
{\plotoneNew{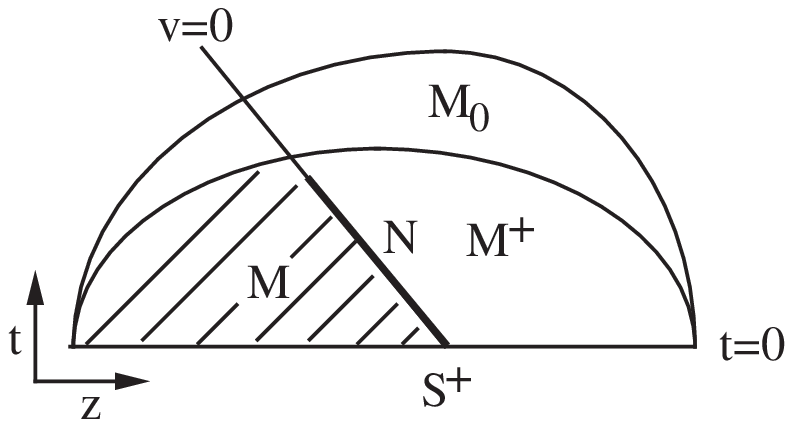}}
{\vskip -0.2truein}
\figure{
Spacetime diagram in $z-t$ coordinates, illustrating the 
mathematical construction used.  Our final spacetime $(M,g)$ consists
of the shaded region.
\label{fig1}}
{\vskip 0.6cm}

We conclude that there exists a class of solutions to
Eq.~(\ref{scalareqn}), which depends on two analytic functions of
($x,y,z$)  
($f_1$ and $f_2$) that can be chosen arbitrarily (apart from 
the above inequality), and which contains a singularity on a 
null hypersurface. In other words, the scalar field admits a 
functionally generic null singularity. (Note that $\phi$ has a well-
defined limit on the singular hypersurface; this is the 
scalar-field analog of the notion of weak singularity.) 

We turn now to generalize this construction to the 
gravitational field. As before, our coordinates are denoted 
($x,y,u,v$).  We adopt the gauge
\begin{equation}
g_{ux}=g_{uy}=g_{uu}=g_{vv}=0, 
\label{eqn110}
\end{equation}
which in turn implies that $g^{vv}=0$.  This ensures that the
coordinate $v$ is null (that is, the hypersurfaces $v=const$ are
null). There are six non-trivial metric functions, which we denote by
$g_i\;(i=1, \cdots, 6)$, where here and below the indices $i,j, \ldots$
represent the six pairs of spacetime coordinates
$(xx,xy,yy,vx,vy,uv)$.

In this gauge, the number $k$ of arbitrary functions in a general
solution is $k=8$.  This can be seen as follows.  Define the new
variables $T\equiv v+u\;,\; Z\equiv v-u$.  Then to determine a
solution of the evolution equations, twelve initial functions need to
be specified on the spacelike hypersurface $T=const$, namely
$g_i(x,y,Z)$ and $g_{i,T}(x,y,Z)$, $1\le i \le 6$.
However, these $12$ functions must satisfy $4$ constraint equations,
as is always the case in general relativity, so that the number of
independently specifiable functions is $k=8$. 
This conclusion can also be reached by adding the conventional number
of intrinsic degrees of freedom of the vacuum gravitational field
($2\times2=4$) to the number of unfixed gauge degrees of freedom in 
the gauge
(\ref{eqn110}), which we show in Ref.~\cite{Flanagan_Ori} to be $4$.

We shall now outline the generalization of the above scalar-
field construction to the gravitational field. First, one 
writes the Einstein equations $R_{\alpha \beta}=0$ in the 
gauge (\ref{eqn110}).  These equations can be naturally divided into 
six evolution equations and four constraint equations. 
At this stage we focus attention on the evolution equations, 
which  can be taken to be $R_i=0$.  Next, we define $w$, $t$ 
and $z$ as before [Eqs.~(\ref{eqn20}),(\ref{eqn45})], and transform the 
field 
equations from the independent variables ($u,v$) to ($z,t$). 
[To avoid confusion, we emphasize that what we are doing here 
is {\it not} a coordinate transformation: it is just a change 
of independent variables in the differential equations 
$R_{\alpha \beta}=0$; thus, the unknowns in Eq.~(\ref{eqn180}) below 
are still the six metric functions $g_i$, which correspond to 
the coordinates ($x,y,u,v$).] By taking certain linear 
combinations of the equations $R_i=0$, 
it is possible to rewrite the evolution equations in the 
schematic form
\begin{equation}
g_{i,tt}=f_i(g_j\,,\,g_{j,t}\,,\,g_{j,A}\,,\,g_{j,AB}\,,\,g
_{j,At}\,,\,z,t).
\label{eqn180}
\end{equation}
Here, the indices $A,B$
run over the  ``spatial'' variables $x,y,z$. If we impose certain 
inequalities on the initial data [which ensure that in the 
region of interest $\det (g)\cong -1$], then the functions $f_i$ 
are analytic in all their arguments. [The gauge conditions 
(\ref{eqn110}) are crucial in deriving Eq.~(\ref{eqn180}).]

We now consider the evolution of initial data under the 
system (\ref{eqn180}). As before, we take the initial hypersurface to 
be $t=0$. Equation (\ref{eqn180}) requires twelve initial functions
to be specified on this hypersurface: the six functions 
$h_i(x,y,z)\equiv g_i(x,y,z\,;\,t=0)$, and the six functions 
$p_i(x,y,z)\equiv g_{i,t}(x,y,z\,;\,t=0)$. The form of
Eq.~(\ref{eqn180})
is suitable for an application of the Cauchy-
Kowalewski theorem. Thus, defining  $S^+$,  $M^+$ and  $M$ 
as before, and following the arguments above, we arrive at 
the following conclusion: For any choice of the above twelve 
analytic functions $h_i(x,y,z)$ and $p_i(x,y,z)$ on the 
section  $S^+$ of t=0 (subject to certain inequalities), 
there exists an analytic solution $g_i(x,y,z,t)$ to Eq.~(\ref{eqn180})
in  $M$.  Again, returning from the variables ($z,t$) to the original 
independent variables ($u,v$), we find that the metric 
functions $g_i(x,y,u,v)$ are continuous throughout  $M^+$ 
(and in particular at  $v=0$) and, moreover, are analytic 
throughout  $M$.  However, at the hypersurface  $v=0$, 
$g_{i,v}$ typically diverge like $v^{-1+1/ n}$.  As a 
consequence, the Riemann components $R_{avbv}$ generically 
diverge there \cite{Flanagan_Ori}.  Moreover, it can be shown that the 
scalar 
$K\equiv R_{\alpha \beta \gamma \delta }R^{\alpha \beta 
\gamma \delta }$ also generically diverges at  $v=0$ (like 
$v^{-2+1/ n}$).  However, it is easy to check directly that the
singularity is weak.  Thus, focusing attention on the physical 
region  $M$, we find that the solutions constructed in that 
way are absolutely regular inside the region  $M$, but 
develop a null, weak, scalar-curvature singularity on the portion
$v=0$ of its boundary.

The twelve initial functions $h_i(x,y,z)$ and $p_i(x,y,z)$ 
are subject to four constraint equations.  It is therefore natural to 
expect that eight of these 12 initial functions can be 
chosen arbitrarily. This is not trivial to prove 
mathematically, however, especially because the constraint 
equations (expressed in the variables $x,y,z$) are somewhat 
pathological at $z=0$.  After some effort, we found a 
mathematical construction which proves the above statement. 
More specifically, in our mathematical scheme one is free to 
choose the six $h_i(x,y,z)$, $p_{xy}(x,y,z)$, and one other function
$p(x,y,z)$.  We can then show (using the 
Cauchy-Kowalewski theorem) the existence of a solution of the 
constraint equations (in a neighborhood of $z=0$). The 
remaining initial functions $p_i(x,y,z)$ are then determined 
from that solution. The above eight 
analytic functions can be chosen arbitrarily, up 
to some inequalities.

To summarize, our mathematical construction shows the existence of a
class of (local) solutions to the vacuum Einstein equations, 
which contain a
weak scalar-curvature singularity at the null hypersurface $v=0$, and
which depends on $k=8$ analytic functions of ($x,y,z$). Our
construction therefore demonstrates the existence of a functionally
generic null, weak, scalar-curvature singularity.

The main limitation of our construction is its restriction to analytic
initial functions.  We believe that this is merely a technical
limitation of the mathematical theorems used in our proof, and the 
same physical situation (a null weak singularity) will evolve even if 
the initial functions on $S^+$ are not analytic (provided they are 
sufficiently smooth for $v<0$).
At any rate, it is worthwhile to compare the
mathematical status of our generic null weak singularity to that of
the BKL singularity.  To the best of the authors' knowledge, the
existence of even a single inhomogeneous singular vacuum solution of
the BKL type has not yet been proved mathematically - let alone the
generality of this class of singular solutions.  On the other hand we
have demonstrated rigorously the existence of a huge class of exact
solutions containing null weak singularities.

Our results have a simple intuitive interpretation: In {\it linear}
hyperbolic systems, it is well known that weak discontinuities of
various types can freely propagate along characteristic lines. Our
construction demonstrates that Einstein's equations, despite their
nonlinearity, also behave in this way (at least with respect to the type
of weak discontinuity considered here).  This is perhaps contrary to what was
sometimes thought in the past, but is not really surprising, because,
after all, Einstein's equations are quasi-linear. Thus, what we have
shown is the {\it local} consistency of null weak singularities with
Einstein's equations, despite the non-linearity of the latter.  The
important issue of the {\it onset} of the singularity from regular,
asymptotically flat, initial data (e.g.~in gravitational collapse)
still remains open; this issue is addressed (indirectly) by the
nonlinear perturbation analysis of Ref.~\cite{Ori_Kerr}, but the onset
still lacks a rigorous mathematical proof.

Finally, it should be pointed out that the inner-horizon singularity
that is suggested by perturbation analyses in a
realistic rotating or charged black hole (see,
e.g.~Ref.~\cite{Ori_Kerr}), is qualitatively similar to 
the singularity constructed here, in that it is null and weak.  There
are some important differences, however.  The main difference is that
the structure of the inner-horizon singularity is analogous to what
would have been obtained from our construction if we had set $w=1/\ln|v|$.
Our method of proof does not generalize straightforwardly to this case,
however, because $v$ is no longer analytic in $w$ at $w=0$ (though it
is still $C^\infty$) \cite{nonaxial}.  We hope to discuss the analytic
features of this more realistic null weak singularity elsewhere.

This research was supported in part by the Israel Science 
Foundation administrated by the Israel Academy of Sciences 
and Humanities, by the Fund for Promotion of Research in 
the Technion, and by the US National Science Foundation grants
AST 9114925 and PHY 9220644.



\begin{references}
\bibitem{caveat2}
The arguments of Ref.~\cite{Yurtsever} are based on an example of
two colliding plane waves.  Note, however, that despite what was
argued in Ref.~\cite{Yurtsever}, the existence of a generic null weak
singularity at the inner horizon is absolutely consistent with the
model of two colliding plane waves \cite{Ori_planewaves}.
\bibitem{note2}
This is a slight abuse of the more usual terminology, 
according to which the number of degrees of freedom is said 
to be $k/2$, rather than $k$.
\bibitem{note5}
Strictly speaking, in order for this definition to be useful, the
nature of the dependence of the solutions on the $k$ functions must be
restricted in some (very natural) ways; see Ref.~\cite{Flanagan_Ori}
for further discussion.
\bibitem{nonaxial}
If the perturbation is not axially-symmetric, the singularity structure 
is even more complicated, as the metric perturbation also oscillates in 
$\ln |v|$ ; See Ref.~\cite{Ori_Kerr} (and recall that the coordinate 
analogous to our $v$ is denoted $V$ there).
\end{references}
\end{document}